# Synthesis of large-area rhombohedral few-layer graphene by chemical vapor deposition on copper

Chamseddine Bouhafs, Sergio Pezzini, Neeraj Mishra, Vaidotas Mišeikis, Yuran Niu, Claudia Struzzi, Alexei A. Zakharov, Stiven Forti, Camilla Coletti*

Dr. C. Bouhafs, Dr. S. Pezzini, Dr. N. Mishra, Dr. V. Mišeikis, Dr. S. Forti and Dr. C. Coletti
Center for Nanotechnology Innovation @ NEST
Istituto Italiano di Tecnologia
Piazza San Silvestro 12, 56127 Pisa, Italy
E-mail: camilla.coletti@iit.it

Dr. C. Bouhafs, Dr. S. Pezzini, Dr. N. Mishra, Dr. V. Mišeikis and Dr. C. Coletti
Graphene Labs
Istituto Italiano di Tecnologia
Via Morego 30, 16163 Genova, Italy

Dr. Y. Niu, Dr. C. Struzzi and Dr. A. A Zakharov
MAX IV Laboratory, University of Lund, 22100, Lund, Sweden

Rhombohedral-stacked few-layer graphene (FLG) has been receiving an ever-increasing attention owing to its peculiar electronic properties that could lead to enticing phenomena such as superconductivity and magnetic ordering. Up to now, experimental studies on such material have been mainly limited by the difficulty in isolating it in thickness exceeding 3 atomic layers with device-compatible size. In this work, rhombohedral graphene with thickness up to 9 layers and areas up to ~50 μm$^2$ is grown via chemical vapor deposition (CVD) on suspended Cu foils and transferred onto target substrates via etch-free delamination. The domains of rhombohedral FLG are identified by Raman spectroscopy and are found to alternate with domains of Bernal-stacked FLG within the same crystal in a stripe-like configuration. A combined analysis of micro-Raman mapping, atomic force microscopy and optical microscopy indicates that the formation of rhombohedral-stacked FLG is strongly correlated to the copper substrate morphology. Cu step bunching results in bending of FLG and interlayer displacement along preferential crystallographic orientations, as determined experimentally by electron microscopy, thus inducing the stripe-like domains. The growth and transfer of rhombohedral FLG with the reported thickness and size shall facilitate the observation of predicted unconventional physics and ultimately add to its technological relevance.

**1. Introduction**

The number of atomic layers and the interlayer stacking order determine the physical properties of van der Waals materials (vdWm) [1]. Few-layer graphene (FLG) is a naturally occurring vdWm, comprising graphene sheets in number $N_G \geq 3$, with two stable configurations, characterized by either Bernal (ABA) [2], or rhombohedral (ABC) [3] stacking. In ABA stacking, within the $N_G = 3$ minimal constituent, the positions of the atoms of the topmost layer exactly match those of the bottom layer. In ABC stacking each layer is laterally shifted with respect to the layer below by an interatomic spacing distance. With respect to the electronic properties, ABA-FLG is a semi-metal with overlapping bands. [4,5] In contrast, ABC-FLG possesses a tunable bandgap [6–8] and surface states characterized by a flat band, the extent of which in reciprocal space enlarges at the increase of the number of layers $N_G$. [9–13] Due to the large density of states and reduced kinetic energy within the flat band, ABC-FLG is susceptible to strong electronic correlation, which makes it a model system for investigating high-temperature superconductivity [11,14,15] and magnetism, [16–19] as well as to realize novel electronic applications. [20]

The main factor constraining the flourishing of ABC-FLG in such research fields is the lack of a synthesis and/or transfer techniques that can yield high-quality large-area FLG with controllable properties ($N_G$ and stacking order) on insulating substrates. Micro-mechanical exfoliation can produce pristine ABC-FLG domains with different thicknesses ($N_G$ from 3 up to 27) and relatively large lateral size (up to few tens of micrometers). [13,20–22] Kish graphite, as the material source for exfoliation, contains typically 80% Bernal, 14% ABC, and 6% turbostratic structure. [3,22] As a result, the exfoliation technique suffers from limited yield of ABC-FLG. Although suitable for fundamental studies, this method is not scalable and the control over $N_G$ remains approximate and not reproducible. ABC-FLG with high crystal quality has been synthesized using thermal decomposition (i.e. sublimation) of silicon carbide (SiC), both on 3C-SiC(111) [12,23,24] and 6H-SiC(0001). [24] However, the size of the domains with uniform thickness presently remains limited to few hundred nanometers and it is generally not sufficient for device fabrication. [24,25]

Furthermore, ABC-FLG is subjected to phase transformation to ABA under external stimuli, such as shear stress, [21] joule heating and laser illumination [26] or field-effect doping. [26] Therefore, alternative methods for the production of ABC-FLG over large scale need to combine with clean transfer and handling approaches capable of preserving the rhombohedral stacking.

Chemical vapor deposition (CVD) is considered a viable route to synthesize high-quality large-area single-crystal monolayer (1L) graphene on copper (Cu) substrates. [26–30] However, the growth of large-area FLG with controlled $N_G$ and stacking using CVD on Cu is challenging, primarily due to a self-limiting mechanism after the formation of 1L. [31] Recent progresses in CVD growth showed the possibility to controllably synthesize large-area bilayer (2L) and trilayer (3L) graphene with different stackings. For example, several hundreds of micrometers large AB-stacked 2L graphene was grown on Cu foils using the so-called pocket growth technique. [28] Centimeter-scale ABA-stacked 3L graphene films were obtained by CVD on single-crystal Cu/Ni(111) alloy.[32] Nevertheless, ABC-3L graphene remains limited to ~$10^2$ nm wide domains when synthesized on Cu [33] or Cu/Ni substrates. [34] Taming CVD growth to obtain high-quality large-area of ABC-FLG with $N_G > 3$ would facilitate further understanding of the physical properties of ABC-FLG and its integration in novel device concepts.

In this work, we have used low-pressure chemical vapor deposition (LPCVD) to grow FLG on Cu substrate. By optimizing our growth condition, we have obtained large-area FLG with varying $N_G$ (up 9L). Raman investigation of the FLG after transfer to $SiO_2$ shows that they are composed of alternating domains with ABA and ABC stacking. The domains have a stripe-like geometry and unpreceded large sizes for CVD-derived ABC-FLG (up to 4 μm width and $10^2$ μm long) are detected. Combined experimental observations reveal how the Cu surface morphology controls the occurrence of the ABC stripes.

## 2. Results and Discussion

As shown in **Figure 1**(a) and (b), we developed a new layout for the growth on Cu foil. This configuration combines the advantages of the back-side diffusion process of Cu pockets [28] with the ability of keeping a flat growth template (see Figure 1(c)), which gives considerable advantages for the subsequent transfer. The Cu foil is suspended over two Cu supports (about 100 μm thick) placed on top of the graphite susceptor. While the top face of the foil, directly exposed to the gas flow, undergoes full coverage of graphitic material, we successfully grew isolated FLG crystals on the bottom side. Figure 1d shows a typical dark field-optical (DF-OM) image obtained after growing FLG single-crystals on the back–side of a suspended Cu foil (Figure 1c). Numerous 1L crystals are clearly visible, with distinctive hexagonal shape and partial merging between adjacent ones. The typical lateral size of the 1L crystals is 600 μm. FLG areas, positioned at the center of the 1L, are detected as brighter regions with typical 90 μm lateral size. In general, the visibility of the 1L crystals and of the adlayers in DF-OM is different for different Cu grains (see full foil DF-OM in Figure S1, Supporting Information). A zoom-in view of FLG crystal F1 (red rectangle in Figure 1d) is shown in Figure 1e. As already reported in other works, [35–37] during CVD growth the Cu surface underneath graphene undergoes step bunching (SB). The step properties height depends on $N_G$ and it is found that increasing $N_G$ favors higher steps. [35,36] SB with different morphology Rayleigh scatter the incident light with varying intensities, resulting in different DF-OM response, with higher steps being brigther. [35] In Figure 1e one can appreciate the wavy-like Cu steps underneath graphene, as well as their increasing brightness as a function of $N_G$, which permits a conclusive distinction of adlayers up to 3L.

To gain further information about $N_G$ and the orientation of the adlayers in the FLG crystals, the as-grown samples are transferred on $SiO_2$/Si substrates (Figure 1f) with a semi-dry transfer approach, [38] that allows to maintain the exceptional properties of graphene as recently demonstrated by transport measurements on 1L. [39] We noticed that transferring FLG results in a higher number of tears and breaks (mostly observed on the 1L parts) with respect to transferring 1L or 2L crystals with the same technique. This is likely due to the fact that thick FLG cause a larger amount of stress on the 1L, which acts as a support layer for FLG during detachment from Cu. Nonetheless, the majority of the transferred FLG were intact in the regions with $N_G \geq 3$, which represent an ideal platform for experimental studies. Figure 1g shows a brigth filed- optical imege (BF-OM) of the same FLG shown in DF-OM in Figure 1e (F1), after semi-dry transfer. On $SiO_2$/Si, $N_G$ can be straightforwardly identified [40]: up to 8L thickness is observed. From BF-OM after transfer, we find that the typical domain size of the FLG grown with our process are approximately 500 μm, 150 μm, 90 μm, 70 μm, 50 μm, 35 μm, 25 μm, 20 μm and 7 μm for 1L, 2L, 3L, 4L, 5L, 6L, 7L, 8L and 9L, respectively. Additionally, BF-OM allows for a straightforward discerning of the relative interlayer rotation in CVD-grown FLG crystals, which is revealed by the orientation of the hexagonally-shaped adlayers. We find that, for our growth conditions, 40% of the FLG crystals present aligned stacking (the inner and outer hexagons have the same orientation), while the remaining 60% show rotated R30 or turbostratic stacking [41]. Such optical assessment was confirmed by Raman analysis, based on the 2D band lineshape and width [42,43] (see Figure S2 in Supporting Information, for a comparison between the Raman spectra of 2L and 3L with aligned and turbostratic stackings).

To check for the presence and properties of ABC-FLG, we carried out a detailed Raman study on the FLG crystals that showed aligned stacking (i.e. no interlayer rotation) in BF-OM. To date, the Raman signatures of ABA and ABC-stacking in exfoliated FLG have been widely investigated. It has been shown that ABA and ABC-FLG have a distinct Raman 2D band lineshapes [22,44–46]. In comparison to ABA-FLG, the 2D band of ABC-FLG is more asymmetric and broader, with a shoulder located on the lower-frequency side [22]. As a consequence, the full-width-at-half-maximum of the 2D band (FWHM(2D)) of ABC-FLG is larger than that of ABA-FLG. Hence, the spatial distribution of ABA and ABC-stacked domains within our FLG can be visualized by plotting FWHM(2D) obtained from Raman mapping,

In **Figure 2a** we show a BF-OM image of large FLG crystals (F2) on $SiO_2$/Si, with thickness up to 9L (see Figure S3 in the SI). Figure 2b shows a color map of FWHM(2D), acquired over the area indicated by the black rectangle in panel (a). Analogous BF-OM and Raman mapping over other two FLG crystals (F3 and F4) are shown in Figure S4, Supporting Information. The Raman 2D band is fitted using a single Voigt function, which gives a reliable estimate of FWHM(2D). FLG exhibits two distinctive type of domains, with clear difference in contrast for fixed $N_G$. The domains form parallel stripes, showing alternating narrow and broad FWHM(2D), and extending across regions with different $N_G$. To understand whether these features correspond to different stacking orders, we have investigated in more detail their Raman 2D band. In Figure 2c we plot representative 2D bands for the two type of domains, for 3L, 4L, 5L, 6L and 7L graphene. Raman data from the stripes with low and high FWHM(2D) observed in Figure 2b are plotted as blue and red lines, respectively. We found that the two of domains show distinctive Raman 2D bands. The lineshape for the areas with low and high FWHM(2D) are entirely comparable to those reported for exfoliated FLG with ABA and ABC stacking, respectively [22,44,46].

To further support our observations, we investigate the Raman M band of each stripe-domain. The M band is extremely sensitive to $N_G$ and FLG stacking order, it consists of low-intensity Raman peaks and it is located in the range 1650 to 1800 cm$^{-1}$ [46–48]. The peaks include a LO+ZA combination mode (at ~1655 cm$^{-1}$), LO+ZO' (~1748 cm$^{-1}$) combination modes and an overtone mode 2ZO (~1775 cm$^{-1}$). The LO and ZO' are in-plane and out-of-plane interlayer breathing optical phonon modes, respectively.

ZA (ZO) is the out-of-plane acoustic (optical) mode in FLG. It was found that LO+ZA and 2ZO do not depend on $N_G$ and the stacking order. [46,47] However, for ABA-FLG, LO+ZO' consists of an asymmetric enhanced peak which blueshifts with increasing $N_G$, while its shape remains almost unchanged. For ABC-FLG, LO+ZO' splits into sharp subpeaks and the number of subpeaks increases with $N_G$. Hence, LO+ZO' can be used to identify ABA and ABC domains in FLG. [46,47] Figure 2d shows the Raman spectra in the M band region for our 3L, 4L, 5L, 6L and 7L graphene. In ABA-FLG (blue lines), LO+ZO' consists of one asymmetrical peak centered at ~1745 cm$^{-1}$ (referred here as $P_1$) and the peak energy increases with $N_G$ (up to 5L). The LO+ZO' splits into two subpeaks ($P_{21}$ and $P_{22}$) for ABC 3L and 4L. For ABC 5L, 6L and 7L, LO+ZO' splits into three ($P_{21}$, $P_{22}$ and $P_{23}$), four ($P_{21}$, $P_{22}$, $P_{23}$ and $P_{24}$) and five ($P_{21}$, $P_{22}$, $P_{23}$, $P_{24}$ and $P_{25}$) subpeaks, respectively. Our results for the M band are in excellent agreements with those reported for exfoliated ABA and ABC-FLG and conclusively identify our domains with different FWHM(2D) (Figure 2b and Figure S4 in Supporting Information) as ABA and ABC-FLG. [46,47]

Via a statistical analysis of our samples, we have found that FLG-crystals with aligned adlayers present either alternating stacking (i.e. a coexisting ABC and ABA domains within the same FLG crystal) or purely ABA stacking. The typical width of the ABC stripes is about 1-4 µm while their length extends up to a few tens of micrometers (for example, in Figure S4d in Supporting Information, the ABC stripes of 5L are 3 µm wide and 20 µm long). It should be mentioned that in all the samples analyzed, we found FLG crystals with alternating ABA and ABC stacking. Over the sample shown in Figure 1f, we have examined 10 aligned FLG crystals, out of which 4 contained FLG with alternating stacking order (40 %).

The distribution of the ABA and ABC domains in our FLG crystals resembles the SB pattern developed by Cu upon graphene growth (appreciable in DF-OM Figure 1e). Indeed, when comparing BF-OM images of the Cu substrate after transfer of FLG with the FWHM(2D) Raman maps of the same FLG crystals on $SiO_2$/Si, the origin of the ABC stripes appears to be indubitably bounded to the Cu substrate morphology. As visible in **Figure 3a** and Figure S5 in Supporting Information, after graphene transfer, the hexagonal shape of the crystals remains imprinted on the Cu foil. The SB associated with different $N_G$ (i.e., increasing step height for increasing $N_G$) is also well visible, allowing one to resolve the growth regions for different FLG thickness. To illustrate the strong correlation between the Cu steps and the domains with different stacking order, in Figure 3b we show a superposition of a BF-OM image of the leftover Cu and a Raman FWHM(2D) map of FLG crystal F3 after transfer (the same is done for F2 and F4 in Figure S5, Supporting Information). The similarity between the shape of the stripe-like domains with alternating stacking and the morphology of the Cu substrate is compelling. The ABA (black-color) and ABC (white-color) regions follow exactly the Cu steps on the growth substrate. In addition, in Figure 3c we show a spatial profile of FWHM(2D) in the 4L region of F3, together with the position of the $P_1$ and $P_{22}$ peaks, which allow for visualizing the alternation between ABC and ABA stripes across the Cu steps. For ABC-FLG, the higher frequency subpeaks ($P_{22}$ in this case) are blueshifted with respect to the main peak $P_1$ of ABA-FLG. Thus, the position of peaks $P_1$ and $P_{22}$ (and $P_{23}$, $P_{24}$ and $P_{25}$ for thicker FLG) tracks the spatial distribution of the ABA and ABC domains identified by the oscillating FWHM(2D).

Figure 3d shows a topographic atomic force microscopy (AFM) image of FLG crystal F1 on Cu substrate (Figure 1e), taken over a 3L to 8L region. Additional AFM topography over 1L-2L areas are shown in Figure S6, Supporting Information. Figure 3e show the height profile along the blue line in Figure 3d. The FLG/Cu surface has a clear terrace-step like structure.

Underneath FLG, the Cu surface morphology changes dramatically. Two parallel alternating regions with different structures and roughness become distinguishable. As shown in Figure 3d and e, the first region (dubbed α) is composed of one large terrace-step, with a step height of about 80-100 nm and with terrace width of 1-2 μm. The other region (dubbed β) is composed of high-density terraces-steps, with ~400 nm wide terraces with step height ~40 nm. By comparing Figure 3e and Figure 3c, one observes that the FLG/Cu corrugations closely resemble the spatial correlation of the Raman parameters in FLG, with the ABA/ABC boundaries corresponding to transitions between α and β type SB.

SEM imaging (Figure S7, Supporting Information) reveal that the surface of the leftover Cu substrate (from crystal F2) is similar to that of FLG-covered Cu. In the region corresponding to $N_G \geq 3$ the two different SB structures (α and β) are also present. This result indicates that the growth-determined Cu surface morphology is preserved after transfer of FLG. More, importantly, there is a striking resemblance in size, shape and "waviness" between the Cu steps imaged by SEM and the ABA and ABC domains revealed by Raman mapping. Figure S8 shows the topographic AFM images of FLG on $SiO_2$ after transfer (crystals F3 and F4). Wrinkles originating from the corrugated structure of FLG/Cu are visible in most cases, confirming that the occurrence of ABC-FLG remains remarkably high after semi-dry transfer.

Having established the correlation between alternating ABA/ABC stripes in CVD-grown FLG and the surface of the Cu catalyst, we now address the mechanisms that underlie the stabilization of the ABC-stacked domains. Gao et al. [34] have recently highlighted the central role of the substrate-FLG interaction and pointed out that the formation of ABC-FLG is promoted by curvature. Despite some similarities with our results, we note that, apart from obtaining smaller (~$10^2$ nm) an thinner (mostly 3L and up to 5L) ABC domains, in Ref.[34] a different catalyst (Cu/Ni alloy), displaying considerable differences in the SB morphology, was used. Therefore, we shall consider several factors specific to our synthesis process.

SB on Cu can be attributed to either strain relaxation during the cool-down step [36,49–51] or minimization of the bending energy of FLG during the growth step, [37] which supports increased-height SB when Cu is covered by thick FLG, as we observe experimentally. Han et al. [52] have recently shown that the bending rigidity of FLG strongly depends on its curvature, reporting considerable softening at large bending angles (>40°). Importantly, this soft regime relies on slipping of atomic planes within FLG, which is proposed as a mechanism stabilizing ABC-FLG, [53] as well as governing FLG stacking transformation under shear stress. [21] We tentatively attribute the formation of ABA/ABC domain walls to interlayer displacements at large bending angle of FLG, in correspondence of the α/β SB boundaries. Nevertheless, according to Refs. [21,53], the direction of interlayer slipping with respect to the FLG crystallographic orientation is crucial in stabilizing ABC-FLG. We shall therefore consider (i) the crystallographic orientation of the hexagonal-shaped FLG and (ii) how the ABC stripes distribute along the Cu steps.

In **Figure 4a,** we show a 25 μm field of view (FOV) LEEM micrograph of FLG F5 with thickness up to 4L, after semi-dry transfer to buffer layer on 6H-SiC(0001) (ZLG/SiC) substrate. Figure 4b shows the corresponding reciprocal space LEED, which allows identifying the crystallographic directions of FLG. The FLG hexagons are zigzag (ZZ) terminated, as preferentially found for isolated graphene crystals on Cu.[54] An evenly-spaced series of aligned wrinkles is visible approximately along the armchair (AC) direction. These features are imprinted from the Cu morphology and likely indicate the main docking orientation of the crystal along the Cu steps, which is known to be isoenergetic for ZZ and AC edges. [55] The Cu steps under FLG, as we have shown by different imaging techniques both before and after transfer of FLG, have a characteristic wavy morphology (see Figure 3d, Figure S5 and Figure S7 in Supporting Information). This implies that bending of FLG and, consequently, interlayer slipping take place along different directions, determined by the wavy SB pattern. As illustrated in the simple sketch of Figure 4d, interlayer displacement along ZZ does not perturb ABA-FLG, while along AC it can drive a stacking transformation to ABC. [21] Indeed, Nery et al. [53] have proposed a strongly orientation-dependent stability diagram for ABC-FLG, which preferentially form under shear-and-slip along AC. We can test this scenario by analyzing a Raman FWHM(2D) map of this specific FLG crystal (F5), as shown in Figure 4c. No ABC domains are found in the left portion of FLG, where wrinkles (derived from Cu steps) are aligned to AC (i.e. the layers displace along ZZ). In contrast, the right part shows numerous ABC stripes, corresponding to in-plane orientation of the Cu steps toward ZZ (interlayer displacement along AC). Assuming ZZ termination for the hexagonal crystals, [54] we can investigate this mechanism also in other FLG samples. In Figure 4e, f and Figure S9 in Supporting Information, we show that the presence of ABC stripes is tied to the local orientation of the wavy Cu steps with respect to the ZZ and AC crystallographic directions. Similarly to the height of the steps, their degree of "waviness" seems depending on $N_G$ rather than being a intrinsic feature of the Cu foil. Our observations suggest a strongly coupled evolution of the FLG/Cu system during growth, which involves and controls the stacking order of FLG.

## 3. Conclusion

In this work, we have introduced a novel approach to the CVD growth of large area FLG crystals, wich can yield ticknesses $N_G$ of 9L or more. Those crystals exhibit alternating stripes with ABA and ABC stacking. ABC-FLG domains show unambiguous Raman fingerprints and extends over device-compatible areas (few tens micrometers long and few micrometers large). Using different experimental techniques, we revealed a strong correlation between the Cu surface morphology after CVD and the formation of ABA/ABC stripes. Our observations can be explained by stabilization of rhombohedral stacking via directional shear-and-slip across boundaries between different SB features on the Cu surface. The synthesis of ABC-stacked FLG domains by LPCVD offers a viable route towards the production of large-scale electronic grade single-crystalline ABC-FLG for applications. Further up-scaling of this approach shall surely rely on deterministic control of the Cu/FLG coupled dynamic along the CVD process.

## 4. Experimental Section

*Growth of FLG on Cu*: FLG samples are grown by using a 4-inch Aixtron BM-Pro cold-wall reactor. An electropolished 25 μm Cu foil (Alfa-Aesar #46365, 99.8 % purity) is used as a substrate and the effective gas flow is reduced by a sample enclosure. [27] The growth process is similar to the one used by Mišeikis et al. [27] and it consists of four steps. The temperature of the furnace is increased up to an optimized value ($T_G$ = 1070 °C, calibrated according to the melting point of Cu) in Ar atmosphere (temperature ramp-up step), then it is maintained constant for 10 min (annealing step). The annealing step serves to reduce the density of nucleation centers. For the growth step, $H_2$ and $CH_4$ (99.99 % purity) are used as effective gases, while Ar is used as a carrier gas. The flow rates are set to 90 sccm, 0.7 sccm and 900 sccm, for $H_2$, $CH_4$ and Ar, respectively. The temperature during growth is kept to $T_G$ ± 1 °C during 90 min. Finally, the chamber is cooled down in Ar and $H_2$ atmosphere (cool-down step). During the whole processes, the pressure is maintained at ~25 mBar.

*FLG transfer to SiO₂/Si and ZLG/SiC substrate:* the as-grown FLG single-crystals on Cu are transferred to either $SiO_2$/Si or buffer layer/6H-SiC(0001) substrates by a semi-dry procedure.[38,56] We use a spin-coated soft polymeric membrane and a semi-rigid PDMS frame to support the crystals during electrochemical delamination in NaOH aqueous solution. The Cu foil is connected to a DC voltage generator and set to 2.4 V with respect to a Pt counter-electrode, during 10-15 min. After detachment, the polymer-supported crystals are released on the final substrate heated at 90°C, the PDMS is peeled off and the membrane is cleaned in organic solvents overnight. This transfer techniques preserve the Cu substrate after the graphene detachment and as result the surface morphology of leftover Cu substrate can be investigated.

*Characterization:* The layer thickness and the stacking order of the resulting FLG crystals on SiO$_2$/Si substrate are determined via micro-Raman scattering spectroscopy. Raman measurements are performed in backscattering geometry under ambient conditions, using a Renishaw InVia spectrometer (1800 grooves/mm) equipped with a Peltier-cooled CCD detector with spectral resolution ~2 cm$^{-1}$. The samples are excited with the 532 nm (2.33 eV) line of a diode-pumped frequency-doubled solid-state Nd: YAG laser through 100× objective (NA 0.9). The laser spot diameter and power are ~1 μm and ~1 mW, respectively. To perform Raman mappings, we used a motorized XYZ stage, with a step size of 0.5 or 1 μm.

The surface morphologies of the FLG samples and Cu foils are characterized using bright-field and optical microscopy, and tapping mode atomic force microscopy. The OM and AFM images are acquired using a Leica DM8000 M light microscope (equipped with a DF condenser) and Bruker Dimension Icon respectively. The morphology of Cu substrates after the graphene transfer is investigated using scanning electron microscopy (SEM). The SEM images are acquired using Zeiss Merlin with an accelerating voltage of 5 kV.

Low- energy electron microscopy (LEEM) and low- energy electron diffraction (LEED) are performed in ultrahigh vacuum and carried out using a SPELEEM instrument at MAXPEEM beamline in the MAX IV Laboratory synchrotron radiation laboratory, Lund, Sweden. Measurements in SPELEEM require conductive and flat substrates. To satisfy these conditions, we have used transferred FLG crystal on buffer layer/6H-SiC substrate (ZLG/SiC). The ZLG/SiC substrate are prepared using sublimation. [57]

**Supporting Information**
Supporting Information is available from the Wiley Online Library or from the author.

**Acknowledgements**

The research leading to these results has received funding from the European Union's Horizon 2020 research and innovation program under grant agreement no. 785219-Graphene Core2.

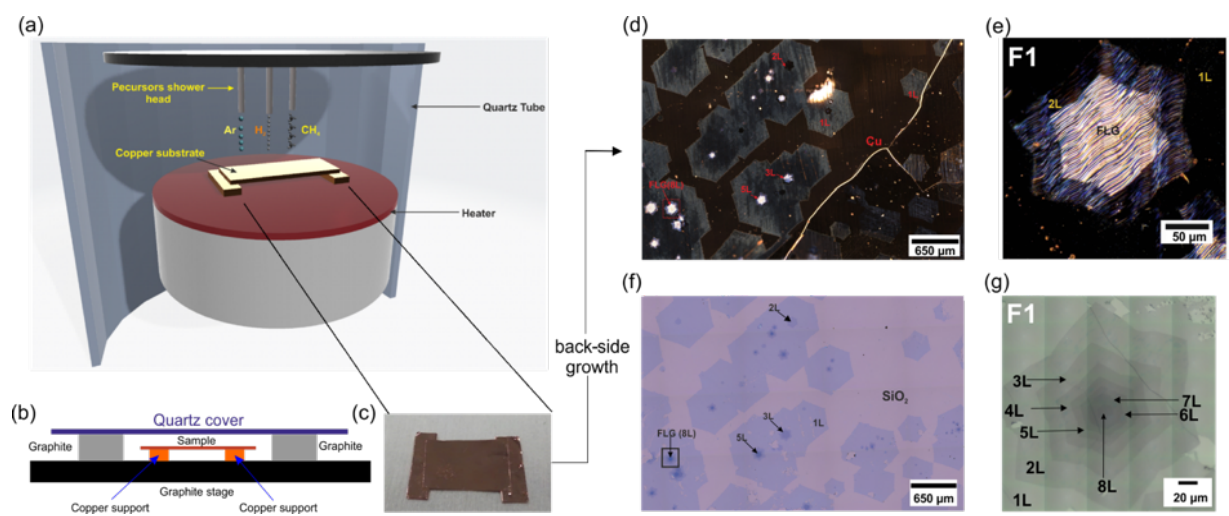

**Figure 1**: a) Schematics of the CVD setup for FLG single-crystals growth. b) Cross sectional sketch of the optimized growth layout. c) BF-OM of a Cu foil suspended over Cu bridges, after FLG growth. d) DF-OM image of a representative as-grown FLG/Cu sample on Cu. e) Zoom-in on FLG crystal F1 (red square in d)). f) BF-OM image of the same FLG sample after transfer to SiO$_2$/Si. g) BF-OM of FLG F1 after transfer (black square in f)).

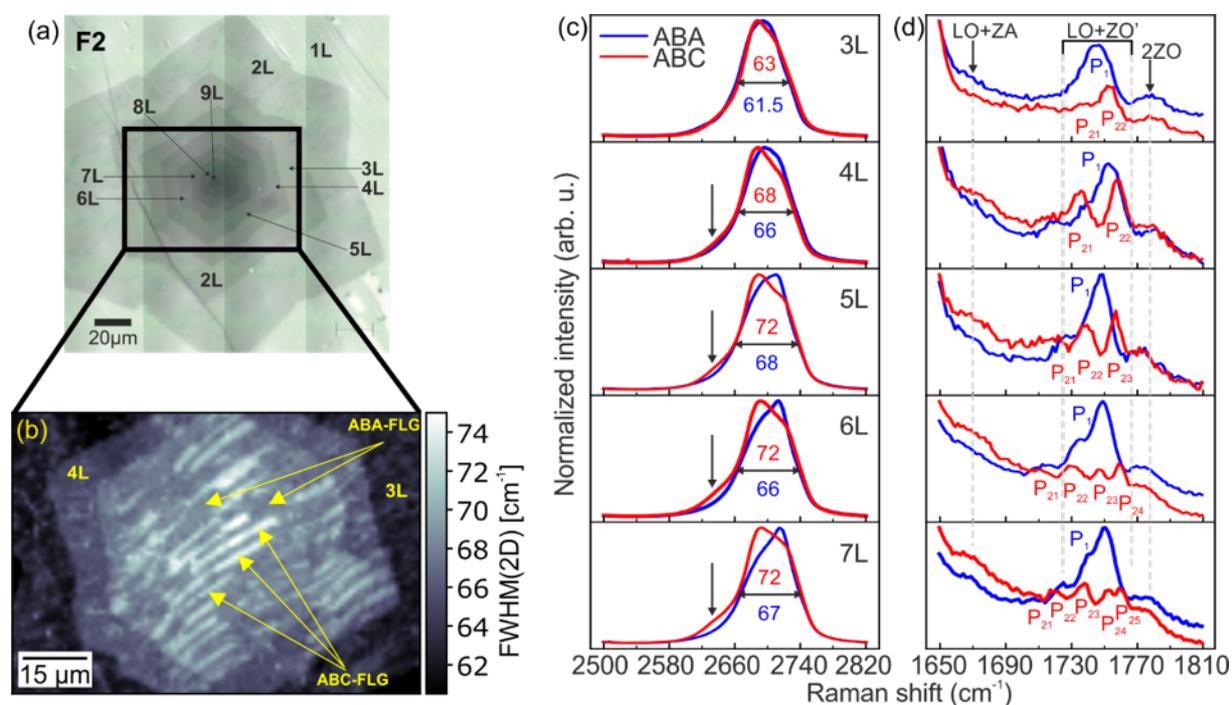

**Figure 2**: a) BF-OM of FLG crystal F2. N$_G$ is indicated and up to 9L can be appreciated. b) FWHM(2D) Raman map, on 3L to 9L FLG area (black rectangle in a); step size is 1 µm. The ABA- and ABC-stacked FLG are indicated. c) 2D band for ABA-stacked (blue) and ABC-stacked (red) FLG, from 3L to 7L. The values of FWHM(2D) are reported and the arrows indicate the enhanced shoulder peak characteristic of ABC-FLG. d) M band for ABA-stacked (blue) and ABC-stacked (red) FLG, from 3L to 7L. The individual components of the M band are indicated.

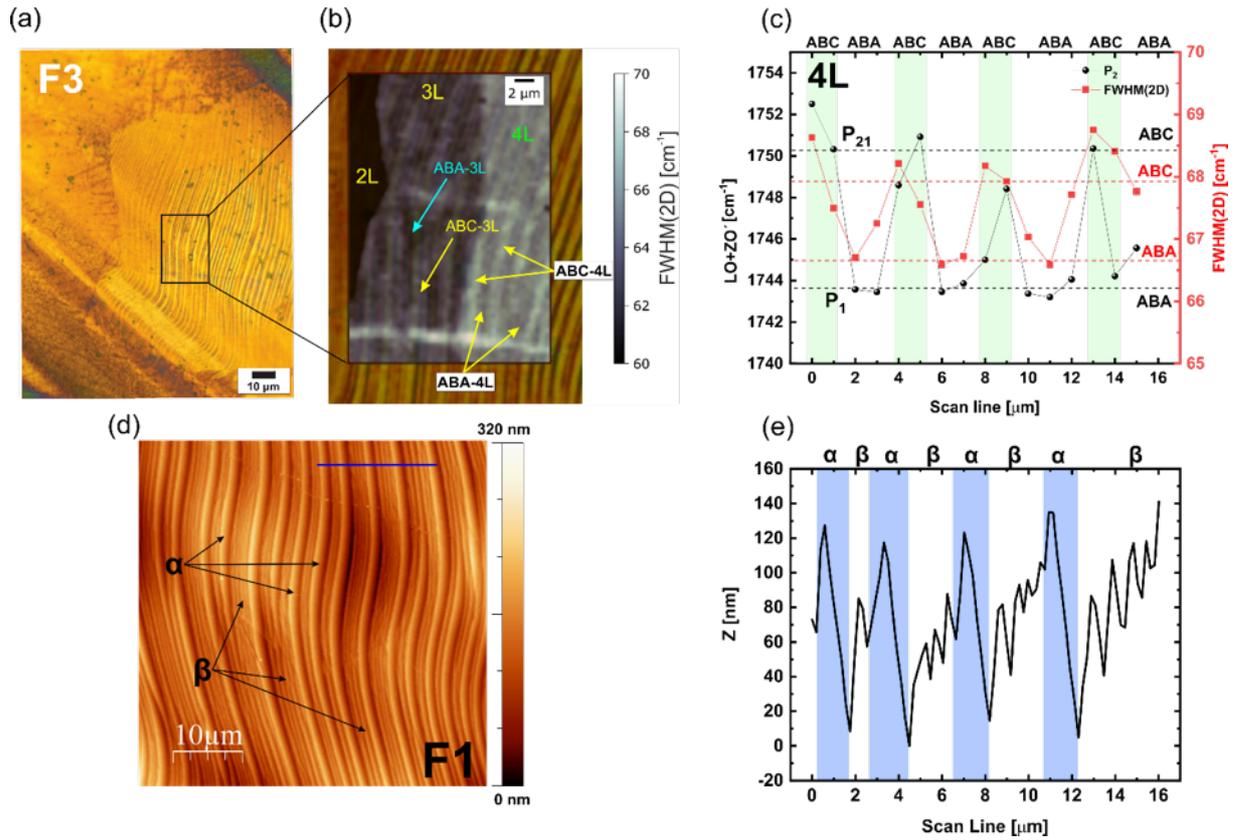

**Figure 3**: a) BF-OM of the Cu substrate after detachment of FLG crystal F3. b) Superposition of a) with the FWHM(2D) Raman mapping of the transferred FLG F3 (see Figure S3c in Supporting Information). c) Spatial correlation of FWHM(2D) and position of peaks P1 and P22, taken along the black line in b). d) AFM image of FLG F1 (3L to 8L thick) on Cu. e) AFM height profile taken along the blue line in d). The x scale matches the one used in c).

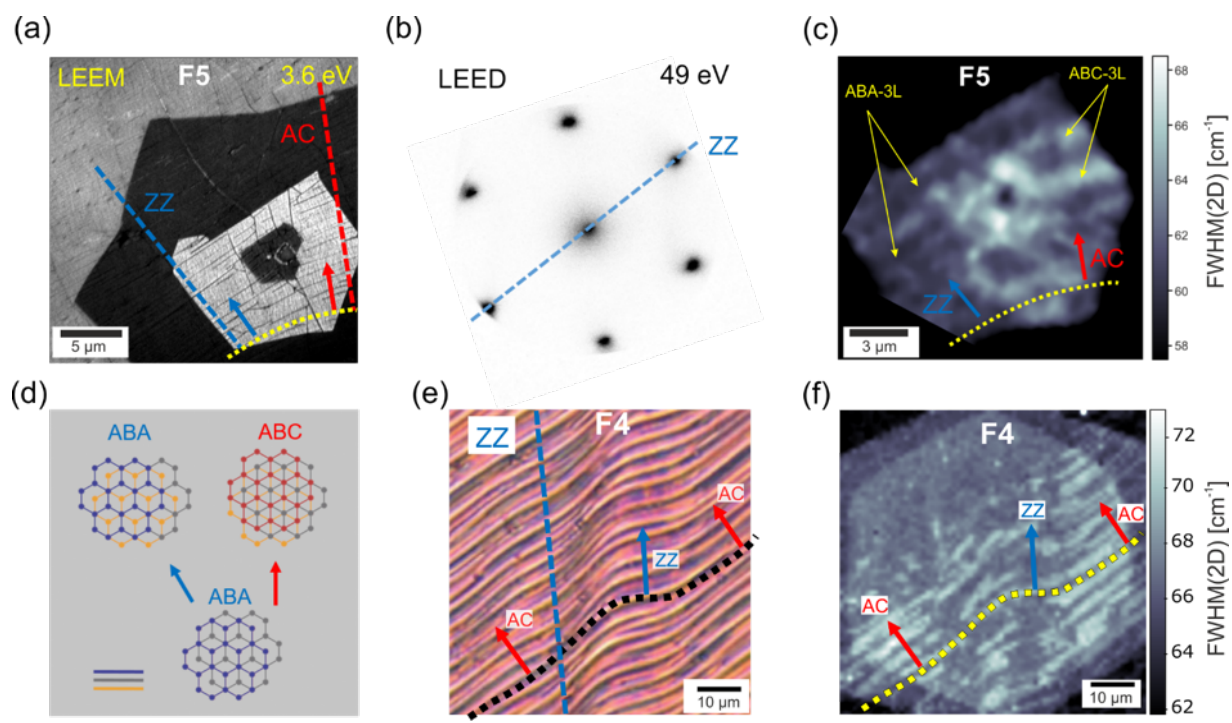

**Figure 4**: a) LEEM real space image of FLG crystal F5 transferred on SiC, recorded at 3.6 eV on a 25 μm FOV. The yellow dotted line traces the imprinted Cu terraces, the blue and red arrows indicate the direction perpendicular to the terraces and approximate the ZZ and AC direction, respectively. b) Reciprocal space LEED image of a), recorded at 49 eV. The blue dashed line identifies the ZZ direction in real space. c) FWHM(2D) map of the FLG in panel a). d) Sketch of the formation ABC staking due to interlayer slipping, limited to a 3L unit. Two of the layers (orange and gray) are kept fixed in the AB configuration, the third one (blue) moves along the ZZ (blue) or AC (red) direction. Similar diagrams are proposed in Ref.[21]. e) BF-OM of Cu foil after detachment of FLG crystal F4. The Cu steps and the slip orientation are indicated as in panel a). f) FWHM(2D) over the FLG isolated from the portion of the foil in e) and transferred to $SiO_2$.

# Supporting Information

Synthesis of large-area rhombohedral-stacked few-layer graphene by chemical vapor deposition on copper

Chamseddine Bouhafs, Sergio. Pezzini, Neeraj Mishra, Vaidotas Mišeikis, Yuran Niu, Claudia Struzzi, Alexei A Zakharov, Steven Forti, Camilla Coletti*

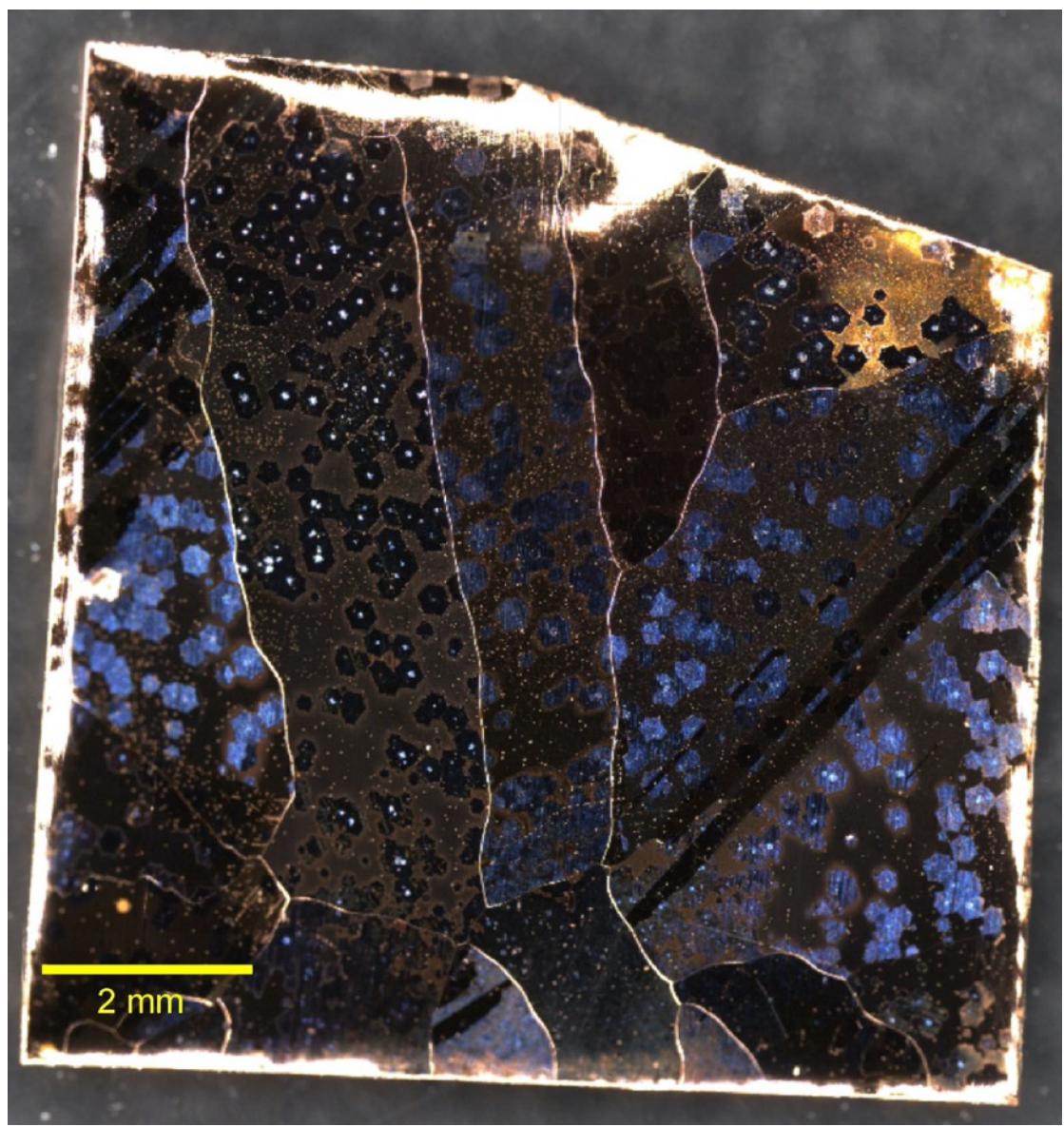

**Figure S1:** DF-OM of Cu foil back-side after FLG growth.

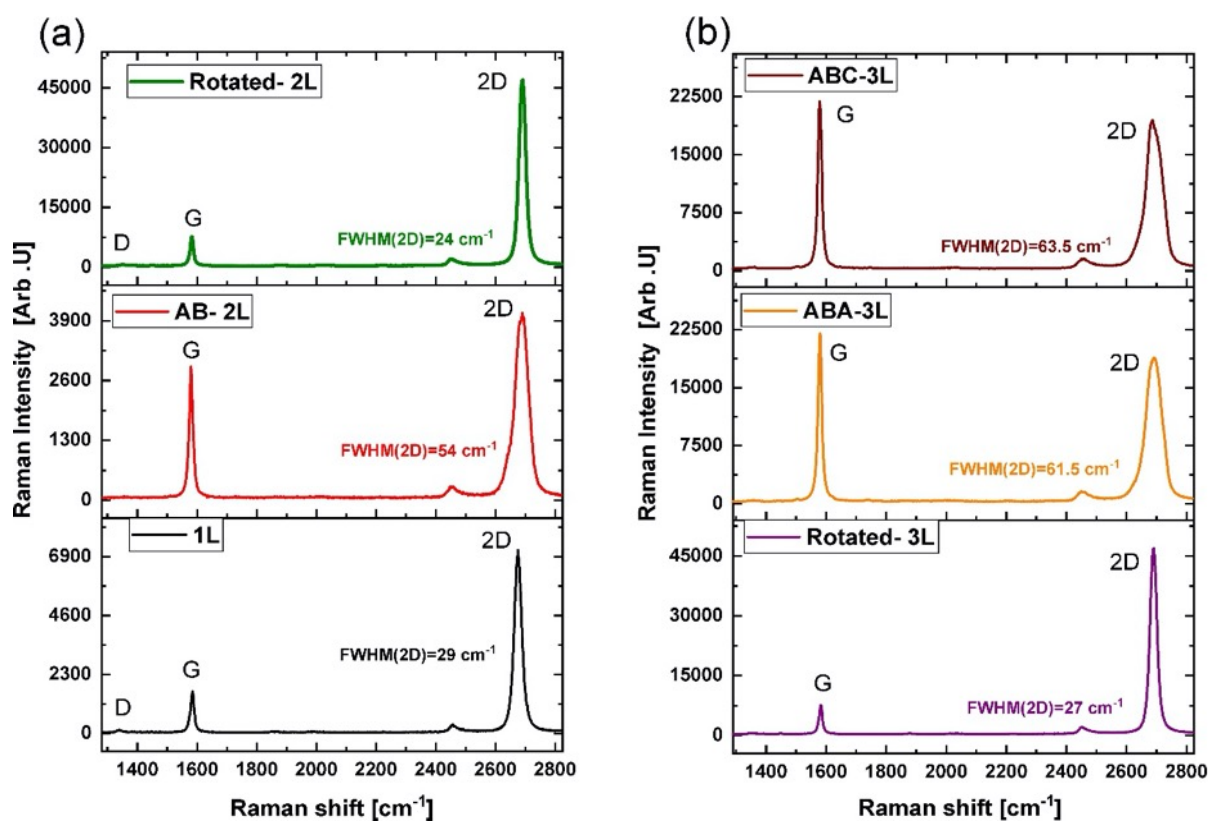

**Figure S2:** a) Raman spectra of 1L (black line), AB- 2L (red) and Rotated- 2L (green). b) Raman spectra of Rotated-3L (violet line), ABA-3L (orange line) and ABC-3L (wine line). In these spectra th the three primary Raman features of graphene are observed. The D band at 1350 cm$^{-1}$ is a defect induce Raman feature. The low intensity D band feature indicates the high crystalline quality of FLG/ SiO$_2$. The G band at 1565-1595 cm$^{-1}$ is associated with the E$_{2g}$ mode at the center of the Brillouin zone and is characteristic of sp$^2$ carbon hybridization. The 2D-band (2600-2800cm$^{-1}$) originates from the two-phonon double resonance Raman process and the lineshape of this band reflects both the phonon dispersion and the electronic band structure. For rotated (Turbostratic) 2L and FLG, the 2D band lineshape is a symmetric Lorentzian similar to 1L graphene. However for AB- 2L, ABC-FLG and ABC-FLG, The 2D band exhibits an asymmetric lineshape and its depend on the number of graphene layers [S1, S2,S3].

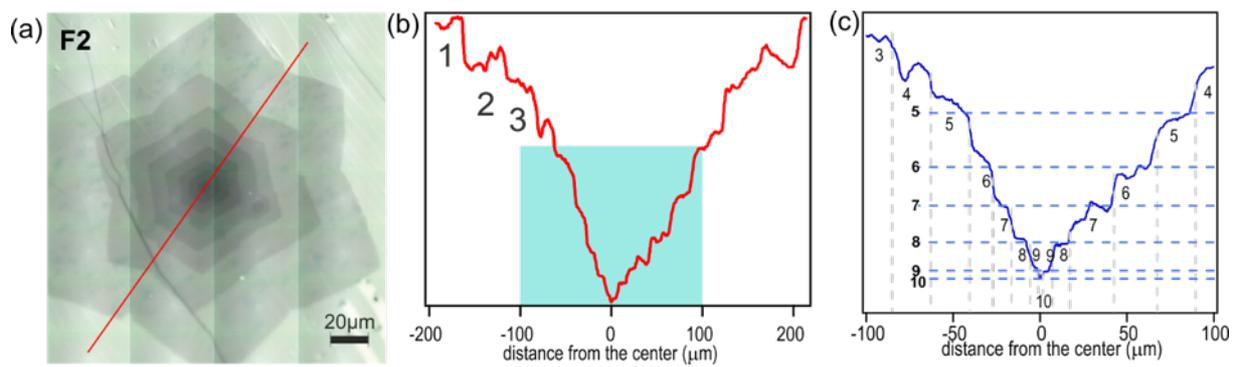

**Figure S3**: a) Optical image of a FLG, as in Fig.2 of the main text. b) Intensity profile recorded along the red line in panel a). c) zoom in within the cyan shaded region in panel b), highlighting the different contrast steps, each corresponding to a different layer thickness.

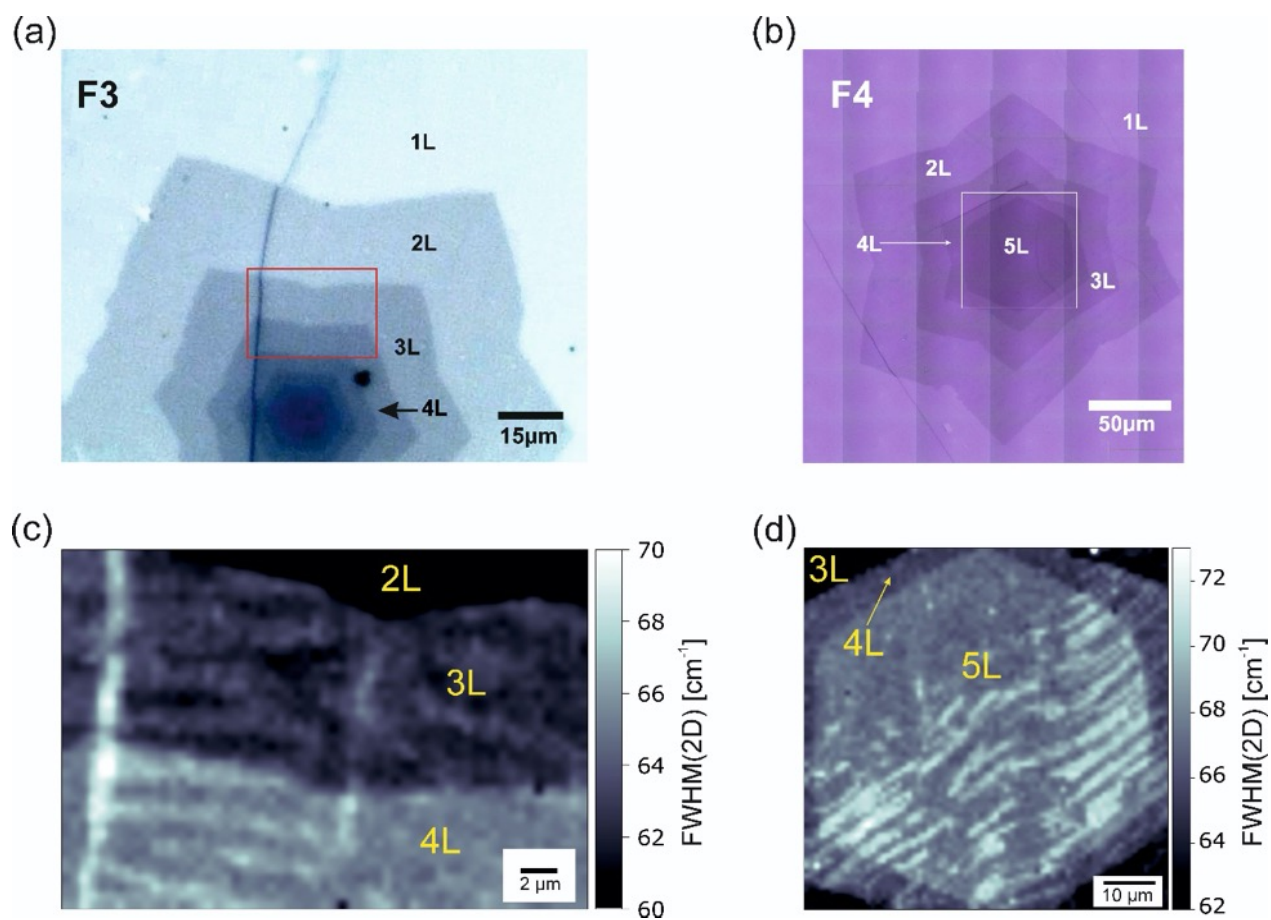

**Figure S4:** a) BF-OM image of FLG crystal F3. b) BF-OM image of FLG crystal F4. c) FWHM(2D) Raman map over the area indicated by the red rectangle in a). d) FWHM(2D) Raman map over the area indicated by the white rectangle in b).

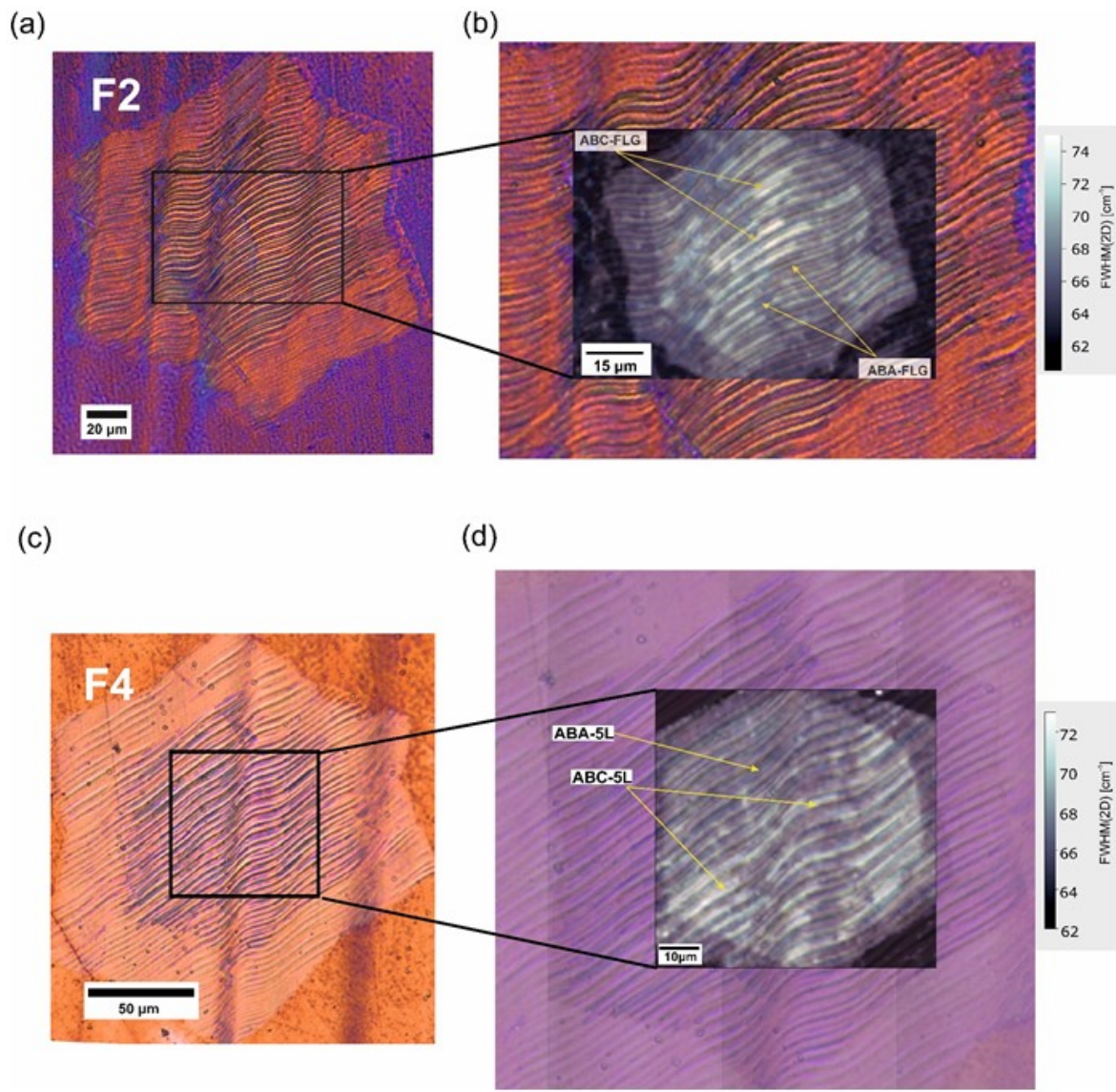

**Figure S5:** a) BF-OM image of Cu substrate after detachment of FLG crystal F2. b) Semi-transparent FWHM(2D) Raman map of F2 after transfer, superimposed to a). c) BF-OM image of Cu substrate after detachment of FLG crystal F4. d) Semi-transparent FWHM(2D) Raman map of F4 after transfer, superimposed to c).

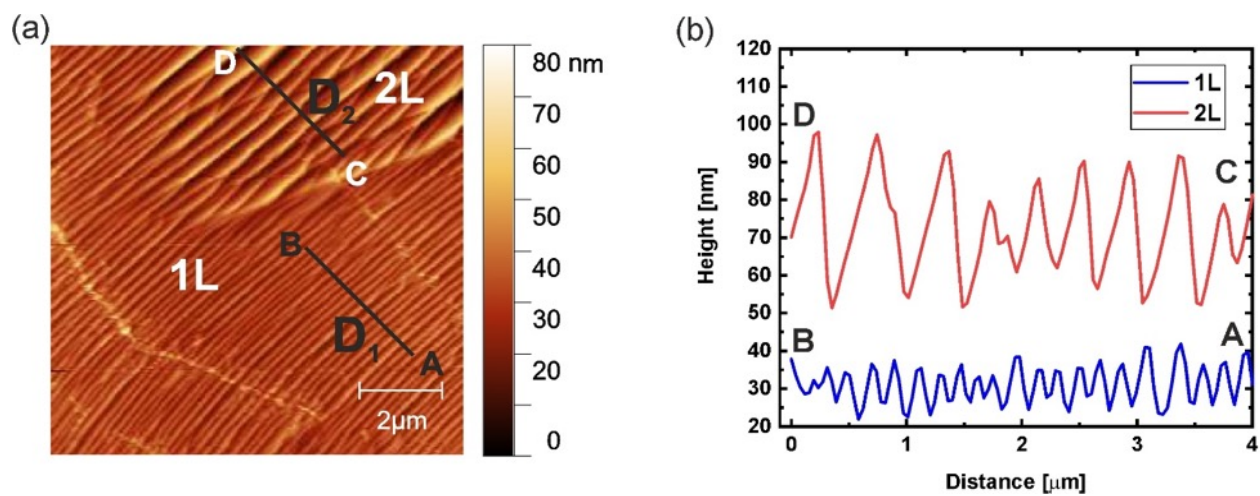

**Figure S6:** a) AFM image of F1 on Cu, over 1L-2L areas. b) AMF height profiles taken along the black lines in a).The Cu surface morphology under 1L is homologous, with narrower features and typical step height 10-15 nm. For 2L graphene, the Cu surface is less regular with higher step height 25-40 nm and wider terraces compared to 1L graphene.

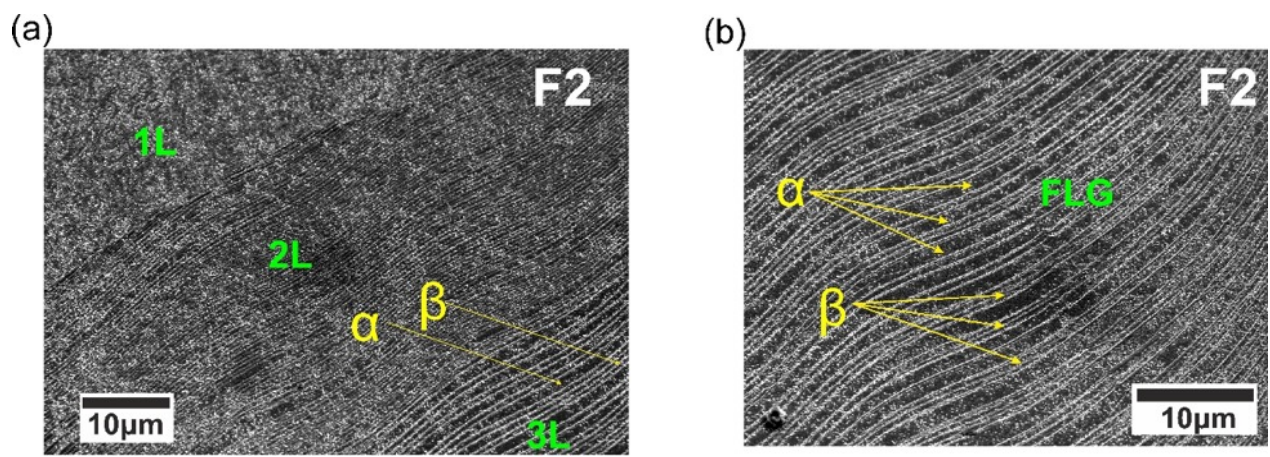

**Figure S7:** SEM images of Cu substrate after detachment of sample F2, acquired over areas corresponding to different $N_G$: a) over 1L, 2L and 3L. b) Over FLG. The two types of SB structures (α and β) are indicated.

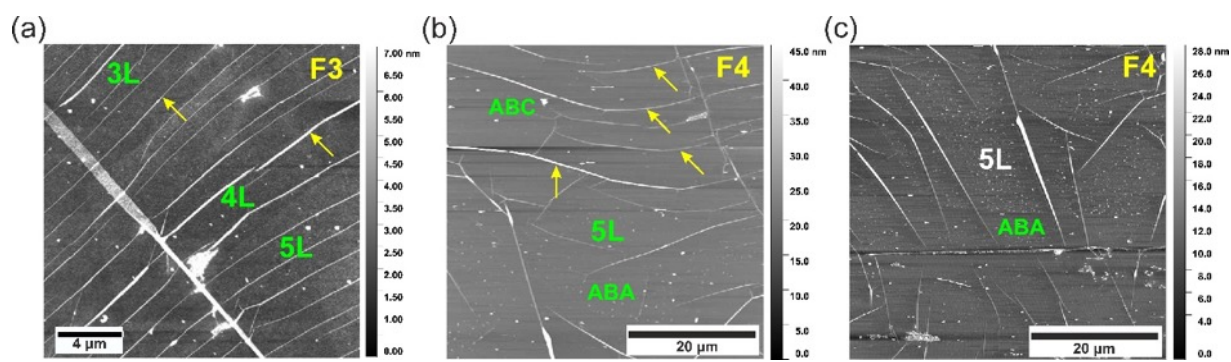

**Figure S8:** AFM images of FLG F3 a) and F4 b),c) on SiO$_2$/Si. N$_G$ and the stacking identified by Raman are indicated. The wrinkles originate from the corrugated structure of FLG/Cu are indicated by yellow arrows. Indeed, they show similar shape as the Cu SB pattern and the alternating domains in Raman mapping, suggesting that the FLG surface morphology and its structure are generally preserved during semi-dry transfer to SiO$_2$ substrate. However, we found that over a few regions the wrinkle network is less ordered and it differs from the Cu morphology (see Figure 4c). In such areas, ABC stacking is not observed by Raman, indicating that stress during the transfer process can induce ABC-ABA stacking transition and domain merging in FLG.

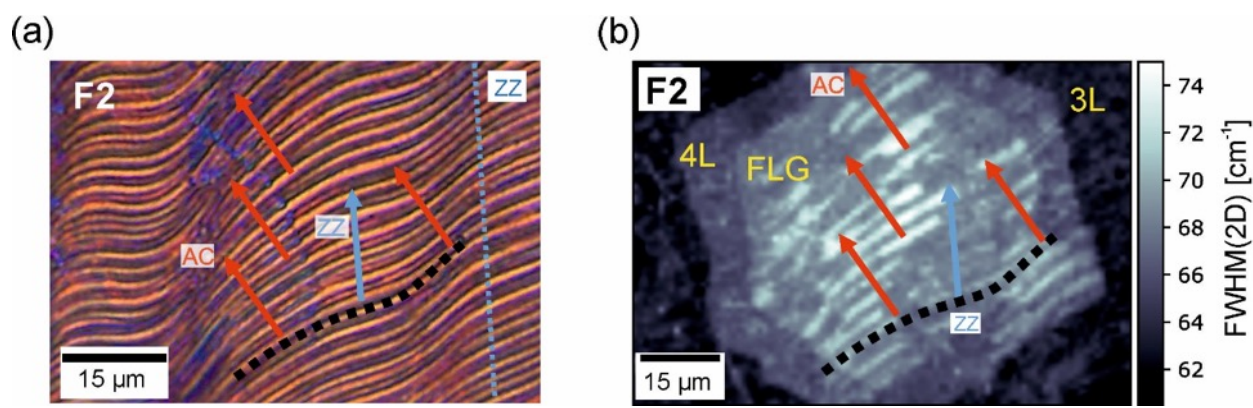

**Figure S9:** a) BF-OM of Cu foil after detachment of FLG crystal F2. The Cu steps and the slip orientation are indicated as in main text Figure 4. f) FWHM(2D) over the FLG isolated from the portion of the foil in e) and transferred to SiO$_2$. The blue and red arrows indicates the ZZ and AC crystallographic orientation, respectively.